\documentclass[aps,pra,showpacs,twocolumn]{revtex4}
\usepackage{graphicx}
\usepackage{bm}
\usepackage{amssymb}

\newcommand{\per}[1]{\mbox{per}\hspace*{1pt}#1\hspace*{1pt}}

\begin{document}
\title{Permanents in linear optical networks}
\author{Stefan Scheel}
\email{s.scheel@imperial.ac.uk}
\affiliation{Quantum Optics and Laser Science, Blackett Laboratory,
Imperial College London, Prince Consort Road, London SW7 2BW, UK}

\begin{abstract}
We develop an abstract look at linear optical networks from the
viewpoint of combinatorics and permanents. In particular we show that
calculation of matrix elements of unitarily transformed photonic
multi-mode states is intimately linked to the computation of
permanents. An implication of this remarkable fact is that all
calculations that are based on evaluating matrix elements are
generically computationally hard. Moreover, quantum mechanics
provides simpler derivations of certain matrix analysis results which
we exemplify by showing that the permanent of any unitary matrix takes
its values across the unit disk in the complex plane.
\end{abstract}

\date{\today}


\maketitle

%%%%%%%%%%%%%%%%%%%%%%%%%%%%%%%%%%%%%%%%%%%%%%%%%%%%%%%%%%%%%%%%%%%%%%
\section{Introduction}

Quantum computing promises to be able to perform certain computational
tasks such as factoring of large numbers \cite{Shor94} exponentially
faster than by using classical computing. Such statements can be made
when looking into computational complexity classes of these
algorithms. Recently, a wealth of proposals for physical
implementation of such algorithms has emerged. The particular one we
have in mind is based on qubit encoding in photons which are subjected
to passive linear transformation in beam splitter networks. These
so-called quantum-gate engineering protocols make use of
measurement-induced nonlinearities in which effective nonlinear
evolutions are achieved by conditional partial measurements in an
extended Hilbert space augmented by auxiliary modes \cite{klm}.
The probabilistic nature of these schemes makes it necessary to
optimize linear-optical networks with respect to their probability of
success. Hence, apart from the complexity of the algorithm that is
supposed to run on the network there is another complexity associated
with the optimization problem.

It seems to have been first noted in \cite{Scheel03} (and somewhat
more discussed in \cite{Schladming}) that optimization of those
networks with respect to their probability of success is closely
linked to maximizing a permanent under certain constraints. 
Permanents have found their way into quantum physics by the work of
Caianiello \cite{Caianiello,CaianielloBuch} who showed that generally
all expectation values of bosonic field operators can be written as a
permanent (or a hafnian, a matrix function related to the permanent
which was actually introduced by him) of a certain coefficient
matrix. At that time it seemed to be little more than a notation
although he was eventually able to sum up Dyson series in field theory
without the use of Feynman graphs. 

In this article we will show that indeed the natural way of writing
unitary transformations of multi-mode Fock states is by means of
permanents of matrices closely associated with the unitary matrix of
the state transformation. This immediately implies that the
calculation of matrix elements of unitarily transformed multi-mode
states (and all subsequent calculations based on it) is
computationally expensive and cannot be performed in polynomial time
(polynomial in the number of modes).

The article is organized as follows. Section~\ref{sec:defs} is used to
introduce some definitions and notations. The main result of this
paper, the unitary transformation of multi-mode states, is then
presented in Sec.~\ref{sec:trafo}. As an example for the application
of these formulas, in Sec.~\ref{sec:power} we describe how the
entanglement power of linear-optical networks can be calculated. The
paper finishes with some concluding remarks in
Sec.~\ref{sec:conclusions}. 

%%%%%%%%%%%%%%%%%%%%%%%%%%%%%%%%%%%%%%%%%%%%%%%%%%%%%%%%%%%%%%%%%%%%%%
\section{Some definitions and notations}
\label{sec:defs}

For the sake of definiteness, we introduce some notation which will be
used throughout this article. To begin with, let us specify what one
commonly understands by the permanent of a matrix. We will concentrate
only on square matrices since this is what we will be using later.

\textit{Def.:} The permanent of an ($n\times n$)-matrix $\bm{\Lambda}$
is a generalized matrix function defined by
\begin{equation}
\label{def:permanent}
\per{\bm{\Lambda}} =  \sum\limits_{\sigma\in S_n}
\prod\limits_{i=1}^n \Lambda_{i\sigma_i} \,,
\end{equation}
where $S_n$ is the group of permutations. For a fixed permutation
$\sigma\in S_n$, the product $\prod\limits_{i=1}^n\Lambda_{i\sigma_i}$
is called a diagonal. The permanent is thus the sum over all diagonals
of the matrix $\bm{\Lambda}$ \cite{Minc}.

It is easy to see that the permanent is similar to the determinant of
a matrix, but lacks the signs related to the order of the
permutation. In fact, these signs cause the permanent to be a very
special matrix function in that all usual computational rules known
for the determinant fail. For example, no rules known from linear
algebra can be used to simplify the computation of the
permanent. Especially, rules such as
$\det\textbf{OSO}^T$ $=\det\textbf{O}\det\textbf{S}\det\textbf{O}^T$
$=\det\textbf{S}$ for orthogonal matrices $\textbf{O}$ do not
hold. The only known simplification for permanents can be achieved for
permutation matrices $\textbf{P}$ and diagonal matrices $\textbf{D}$
in which case we have
\begin{equation}
\per{\textbf{P}\bm{\Lambda}\textbf{D}} = \per{\textbf{P}}
\per{\bm{\Lambda}} \per{\textbf{D}} \,.
\end{equation}
In general, the computation of the permanent has to be done without
possible simplification in which case there are $n!$ terms in the sum
(\ref{def:permanent}). In fact, there exists a result by Marcus and
Minc \cite{MarcusMinc} that asserts that there exists no linear
transformation on $n\times n$-matrices with $n\ge 3$ such that the
permanent of these matrices can be converted into a determinant.

It will prove to be convenient to introduce some more notation. Let
$\bm{\Lambda}[k_1,\ldots,k_m|l_1,\ldots,l_m]$ be the
($m\times m$)-matrix whose matrix elements are those of the original
matrix $\bm{\Lambda}$ with row indices $k_1,\ldots,k_m$ and column
indices $l_1,\ldots,l_m$. For example,
\begin{equation}
\bm{\Lambda}[k_1,k_2,k_3|l_1,l_2,l_3] =
\left( \begin{array}{ccc}
\Lambda_{k_1l_1} & \Lambda_{k_1l_2} & \Lambda_{k_1l_3} \\
\Lambda_{k_2l_1} & \Lambda_{k_2l_2} & \Lambda_{k_2l_3} \\
\Lambda_{k_3l_1} & \Lambda_{k_3l_2} & \Lambda_{k_3l_3}
\end{array} \right) \,.
\end{equation}
The object
$\bm{\Lambda}[(1^{m_1},2^{m_2},\ldots)|(1^{n_1},2^{n_2},\ldots)]$
denotes a matrix whose entries are taken from the matrix
$\bm{\Lambda}$ and whose row index $i$ occurs exactly $m_i$ times and
whose column index $j$ occurs exactly $n_j$ times, for example,
\begin{equation}
\bm{\Lambda}[(1^1,2^1,3^1)|(1^0,2^2,3^1)] = \left(
\begin{array}{ccc}
\Lambda_{12} & \Lambda_{12} & \Lambda_{13} \\
\Lambda_{22} & \Lambda_{22} & \Lambda_{23} \\
\Lambda_{32} & \Lambda_{32} & \Lambda_{33}
\end{array} \right) \,.
\end{equation}

Furthermore, let $G_{n,N}$ be the set of all non-decreasing integer
sequences $\bm{\omega}$,
\begin{equation}
G_{n,N}:= \{ \bm{\omega}:
1\le\omega_1\le\cdots\le\omega_n \le N \} \,. 
\end{equation}
That is, each sequence $\bm{\omega}\in G_{n,N}$ has length $n$ and its
entries take integer values up to $N$.

%%%%%%%%%%%%%%%%%%%%%%%%%%%%%%%%%%%%%%%%%%%%%%%%%%%%%%%%%%%%%%%%%%%%%%
\section{Unitary transformation of multi-mode Fock states}
\label{sec:trafo}

With the above definitions it is now rather straightforward to write
down the expression how a bosonic (photonic) Fock state
$|n_1,n_2,\ldots,n_N\rangle$ transforms under the action of a unitary
network specified by an operator $\hat{U}$ or a unitary matrix
$\bm{\Lambda}$, respectively. Traditionally, this transformation is
written by using the transformation rule for photonic amplitude
operators \cite{VogelWelsch}
\begin{equation}
\hat{\textbf{a}} \mapsto \bm{\Lambda}^+\hat{\textbf{a}} \,,\quad
\hat{\textbf{a}}^\dagger \mapsto \bm{\Lambda}^T\hat{\textbf{a}}^\dagger
\end{equation}
as
\begin{equation}
\hat{U} |n_1,n_2,\ldots,n_N\rangle =
\prod\limits_{i=1}^N \frac{1}{\sqrt{n_i!}} \left(
\sum\limits_{k_i=1}^N  \Lambda_{k_i,i} \hat{a}_{k_i}^\dagger
\right)^{n_i} 
|0\rangle^{\otimes N} \,.
\end{equation}
By using the multinomial expansion theorem, this can be further
rewritten as
\begin{eqnarray}
\label{eq:multinomial_expansion}
\lefteqn{
\hat{U} |n_1,n_2,\ldots,n_N\rangle =
\sum\limits_{ \{n_{ij}\} \atop \sum\limits_{j=1}^N n_{ij}=n_i}
\frac{\prod\limits_{i=1}^N (n_i!)^{1/2}}{\prod\limits_{i,j=1}^N n_{ij}!}
} \nonumber \\ &&
\prod\limits_{j_1=1}^N (\Lambda_{j_11}\hat{a}_{j_1}^\dagger)^{n_{1j_1}}
\ldots
\prod\limits_{j_N=1}^N (\Lambda_{j_NN}\hat{a}_{j_N}^\dagger)^{n_{Nj_N}}
|0\rangle^{\otimes N} \,.
\end{eqnarray}
Note that the $n_{ij}$ form a $n\times n$-matrix whose row sums
are the photon numbers $n_i$, hence $\sum\limits_{j=1}^N n_{ij}=n_i$.
Moreover, the row sums (as well as the column sums to ensure
unitarity, hence conservation of the total photon number) add up to
the total photon number $n$ which makes the matrix $\{n_{ij}/n\}$
(doubly) stochastic. Denoting the column sums of $\{n_{ij}\}$ as
$\sum\limits_{i=1}^N n_{ij}=m_j$, we can write the second line in
Eq.~(\ref{eq:multinomial_expansion}) as
\begin{equation}
\left( \prod\limits_{l=1}^N m_l! \right)^{1/2}
\left( \prod\limits_{k,l=1}^N \Lambda_{lk}^{n_{kl}} \right)
|m_1,m_2,\ldots,m_N\rangle \,. 
\end{equation}
Note that $\prod\limits_{k,l=1}^N \Lambda_{lk}^{n_{kl}}$ is a product
with exactly $n$ factors where each row index $i$ occurs $m_i$ times
and where each column index $j$ occurs $n_j$ times. Therefore, it must
be a diagonal (according to the definition in Sec.~\ref{sec:defs}) of
the matrix
$\bm{\Lambda}[(1^{m_1},\ldots,N^{m_N})|(1^{n_1},\ldots,N^{n_N})]$.
If we take the permanent of this matrix, we see that out of all
possible permutations of column indices, $\prod_j n_j!$ of those
permutations are identical. Analogously, there are $\prod_i m_i!$ ways
of distributing the row indices. Hence, not all diagonals are
different from each other. However, only
$(\prod_i m_i!)(\prod_j n_j!)/(\prod_{i,j} n_{ij}!)$ terms actually
lead to the same diagonal. With the knowledge of these multiplicities
it follows from Eq.~(\ref{eq:multinomial_expansion}) that
\begin{eqnarray}
\label{eq:matrixelement}
\lefteqn{
\langle m_1,m_2,\ldots,m_N|\hat{U}|n_1,n_2,\ldots,n_N\rangle = }
\nonumber \\ &&
\left( \prod\limits_i n_i! \right)^{-1/2}
\left( \prod\limits_j m_j! \right)^{-1/2}
\per{\bm{\Lambda}}[\bm{\Omega}'|\bm{\Omega}]
\end{eqnarray}
where
\begin{eqnarray}
\bm{\Omega} &=& (1^{n_1},2^{n_2},\ldots,N^{n_N}) \,,\\
\bm{\Omega}' &=& (1^{m_1},2^{m_2},\ldots,N^{m_N}) \,.
\end{eqnarray}
Equation~(\ref{eq:matrixelement}) is proof for the intimate relation
between unitary transformation of multi-mode Fock states and
permanents of matrices associated with the unitary matrix $\bm{\Lambda}$.

An immediate consequence of Eq.~(\ref{eq:matrixelement}) is that 
\begin{equation}
\per{\bm{\Lambda}} = {}^{\otimes N}\langle 1|\hat{U}|1
\rangle^{\otimes N}
\end{equation}
with the effect that the permanent of a unitary matrix takes its value
across the unit disk in the complex plane,
$|\per{\bm{\Lambda}}|\le 1$, since the expectation value to find the
state $|1\rangle^{\otimes N}$ after performing a unitary transform of
a similar state is actually a probability amplitude. Although this
result is usually derived from the Marcus--Newman theorem
\cite{MarcusNewman}, the probability interpretation of quantum mechanics
provides a much simpler derivation.

The description (\ref{eq:matrixelement}) of matrix elements can be
straightforwardly generalized to the full unitary transformation as
\begin{eqnarray}
\label{eq:trans}
\lefteqn{
\hat{U}|n_1,n_2,\ldots,n_N\rangle =
\left( \prod\limits_i n_i! \right)^{-1/2}
} \\ && \hspace*{-3ex}
\sum\limits_{\bm{\omega}\in G_{n,N}} \frac{1}{\sqrt{\mu(\bm{\omega})}}
\per{\bm{\Lambda}}[\bm{\omega}|\bm{\Omega}]
|m_1(\bm{\omega}),m_2(\bm{\omega}),\ldots,m_N(\bm{\omega})\rangle
\nonumber 
\end{eqnarray}
in which the $m_i(\bm{\omega})$ are the multiplicities of the
occurrence of the value $i$ in the non-decreasing integer sequence
$\bm{\omega}$ and $\mu(\bm{\omega})=\prod_i m_i(\bm{\omega})!$.

Analogously, we obtain partial matrix elements (or rather projections)
with $\sum_{j=2}^N m_j\le n$ as
\begin{eqnarray}
\lefteqn{
\langle m_2,\ldots,m_N|\hat{U}|n_1,n_2,\ldots,n_N\rangle =}
\\ && \hspace*{-3ex}
\left( \prod\limits_{i=1}^N n_i! \right)^{-1/2}
\left( \prod\limits_{j=2}^N m_j! \right)^{-1/2}
\per{\bm{\Lambda}}[\bm{\Omega}''|\bm{\Omega}]
\left|n-\sum\limits_{j=2}^N m_j \right\rangle \nonumber
\end{eqnarray}
with
\begin{equation}
\bm{\Omega}'' = ( 1^{n-\sum_{j=2}^N m_j},2^{m_2},\ldots,N^{m_N} ) \,.
\end{equation}

Equations~(\ref{eq:matrixelement}) and (\ref{eq:trans}) constitute the
main result of this paper. Either equation implies that quantum-state
transformations can be written in terms of permanents of matrices
associated with the unitary matrix $\bm{\Lambda}$. Although nothing
has changed in the state transformation itself, this result
nevertheless shows that the computational complexity of computing
matrix elements in the Fock basis and functions depending on it is the
same as computing permanents. The intrinsic interest in this result
stems from the fact that it provides an argument for the affiliation
of any type of optimization algorithm for passive linear optical
networks or any maximization or averaging procedure over unitary
operation in the Fock basis to the same complexity class, namely that
of computing a permanent. The following section shall provide an
example for an averaging procedure over unitary operator that can
entirely be written as a function of permanents.

%%%%%%%%%%%%%%%%%%%%%%%%%%%%%%%%%%%%%%%%%%%%%%%%%%%%%%%%%%%%%%%%%%%%%%
\section{Entanglement power of unitary networks}
\label{sec:power}

Over the recent years, entanglement and its quantification has played
a major role in the discussions on the foundations of quantum
mechanics. The basic principle lies in the distinction between
separable and inseparable (entangled) states. Commonly, a bipartite
state is called separable if its density matrix can be decomposed into
a convex sum of tensor product states of the respective subsystems. On
the contrary, an entangled state cannot be written in this form. Apart
from separability criteria that can tell to which of the above classes
a given quantum state belongs, there exist entropic or distance-based
measures to quantify the amount of entanglement. We will not give any
details here since they can be found in the vast literature on this
subject  (for a recent review, see e.g. \cite{entmeas}), we merely
note that for pure bipartite states all of these entanglement measures
collapse to the von Neumann entropy of the reduced density matrix.

A somewhat related question is to ask how much entanglement a quantum
operation can generate when applied to some initial bipartite
state \cite{operations}. Among all possible questions of this
type we will concentrate only on the problem how much entanglement
will be generated on average from a pure product state as opposed to
the maximally available entanglement \cite{maxent}. We will
henceforth call a functional $P_E(\hat{U})$ the entanglement power of
the unitary operator $\hat{U}$ with respect to the entanglement
measure $E$ if
\begin{equation}
P_E(\hat{U}) = \int dg(|\psi_1\rangle) dg(|\psi_2\rangle) \,
E(\hat{U}|\psi_1,\psi_2\rangle) \,,
\end{equation}
where $dg(|\psi_i\rangle)$ denotes the group measure with respect to
the state $|\psi_i\rangle$. In what follows, we will restrict
ourselves to the linear entropy as the entanglement measure of choice,
simply because it is easiest to compute. The linear entropy, defined
by
\begin{equation}
L(\hat{U}|\psi_1,\psi_2\rangle) = 2(1-\mbox{Tr}\hat{\varrho}_1^2)
\end{equation}
where $\hat{\varrho}_1=$
$\mbox{Tr}_2(\hat{U}|\psi_1,\psi_2\rangle\langle\psi_1,\psi_2|\hat{U}^\dagger)$,
although not a proper entanglement measure in the strict sense, serves
as an upper bound to the distillable entanglement.

\subsection{Qubit initial states}

To begin with, let us consider the simplest example in which two qubit
states impinge on a beam splitter with transmittivity
$T\in\mathbb{C}$. We will not use the language of permanents here but
rather perform a straightforward calculation because that appears to
be simpler (we leave the calculation involving permanents to the
example in the next section). Let us define the input states as
\begin{eqnarray}
|\psi_1\rangle &=& c_0 |0\rangle +c_1|1\rangle \,,\\
|\psi_2\rangle &=& d_0 |0\rangle +d_1|1\rangle \,,
\end{eqnarray}
where $|c_0|^2+|c_1|^2=|d_0|^2+|d_1|^2=1$, and let the unitary matrix
associated with the beam splitter be given by
\begin{equation}
\bm{\Lambda} =
\left( \begin{array}{cc}
T & R \\ -R^\ast & T^\ast
\end{array} \right) \,.
\end{equation}
The usual replacement rules for the bosonic amplitude operators,
$\textbf{a}^\dagger\mapsto\bm{\Lambda}^T\textbf{a}^\dagger$,
immediately imply that
\begin{eqnarray}
\label{eq:simple}
\lefteqn{
\hat{U}|\psi_1,\psi_2\rangle = c_0d_0 |00\rangle }
\\ && \hspace*{-3ex}
+(c_0d_1R+c_1d_0T)|10\rangle +(c_0d_1T^\ast-c_1d_0R^\ast)|01\rangle
\nonumber \\ && \hspace*{-3ex}
+c_1d_1 \left[ \sqrt{2}TR|20\rangle -\sqrt{2}T^\ast R^\ast|02\rangle
+(|T|^2-|R|^2)|11\rangle  \right] \,. \nonumber
\end{eqnarray}
The coefficient in the last term in Eq.~(\ref{eq:simple}) is
proportional to the permanent of the beam splitter matrix
$\per{\bm{\Lambda}}=|T|^2-|R|^2$, as one expects from the general
theory presented above (it is just the probability amplitude of
finding the state $|11\rangle$ from a $|11\rangle$ input state). The
reduced density matrix of subsystem 1 is then
\begin{eqnarray}
\lefteqn{
\hat{\varrho}_1 =
-\left[ \sqrt{2} c_1d_1 T^\ast R^\ast |0\rangle \right] \otimes
\mbox{h.c.} 
} \\ && \hspace*{-3ex}
+\left[ (c_0d_1T^\ast-c_1d_0R^\ast)|0\rangle +c_1d_1
(|T|^2-|R|^2)|1\rangle \right] \otimes \mbox{h.c.}
\nonumber \\ && \hspace*{-3ex}
+\left[ c_0d_0 |0\rangle +(c_0d_1R+c_1d_0T)|1\rangle
+\sqrt{2}c_1d_1|2\rangle \right] \otimes \mbox{h.c.} \,.\nonumber
\end{eqnarray}
The remaining steps are to compute the linear entropy and to integrate
over the group measures with respect to the dynamical groups
associated with the initial states $|\psi_i\rangle$. The crucial
observation at this point is that the group integration removes all
phase dependencies from the linear entropy, an effect that can be
attributed to Schur's Lemma. The remaining amplitude dependence
(either on $|T|$ or on $|R|$) can be uniquely rewritten in terms of
the permanent, since $|T|^2=(1+\per{\bm{\Lambda}})/2$ and
$|R|^2=(1-\per{\bm{\Lambda}})/2$. A straightforward calculation then
leads to the result that
\begin{equation}
\label{eq:simplepower}
P_L(\hat{U}) = \frac{3}{64}
\left[ 1-(\per{\bm{\Lambda}})^2 \right]
\left[ 13+9(\per{\bm{\Lambda}})^2 \right]
\end{equation}
which only depends on the permanent of the beam splitter
matrix. Obviously, if the beam splitter acts only as a phase shifter
(in which case $\per{\bm{\Lambda}}=1$) or as a two-sided mirror
($\per{\bm{\Lambda}}=-1$) the input beams do not mix and no
entanglement is created whatsoever.
The maximum of Eq.~(\ref{eq:simplepower}) is reached when
$\per{\bm{\Lambda}}=0$, i.e. when the beam splitter is balanced. The
entanglement power at this point is $39/64$ and monotonically
decreasing with increasing $|\per{\bm{\Lambda}}|$.

\subsection{Higher-dimensional initial states}

Let us now look at the situation in which the initial states that
impinge on a single beam splitter are $N+1$-dimensional, hence can be
written as
\begin{equation}
|\psi_1\rangle = \sum\limits_{n_1=0}^N c_{n_1}|n_1\rangle \,,\quad
|\psi_2\rangle = \sum\limits_{n_2=0}^N d_{n_2}|n_2\rangle \,,
\end{equation}
such that the combined state reads
\begin{equation}
|\psi_1,\psi_2\rangle = \sum\limits_{n_1,n_2=0}^N c_{n_1} d_{n_2}
|n_1,n_2\rangle \,.
\end{equation}
Now we refrain from using the straightforward calculation since that
becomes messy very quickly. Instead, we use the permanent language
developed in the previous section.
Applying the transformation rule (\ref{eq:trans}) to this state yields
\begin{eqnarray}
\lefteqn{
\hat{U} |\psi_1,\psi_2\rangle = \sum\limits_{n_1,n_2=0}^N
\frac{c_{n_1} d_{n_2}}{\sqrt{n_1!n_2!}} 
} \nonumber \\ &&
\sum\limits_{\bm{\omega}\in G_{n_1+n_2,2}}
\frac{\per{\bm{\Lambda}}[\bm{\omega}|(1^{n_1},2^{n_2})]}
{\sqrt{m_1(\bm{\omega})!m_2(\bm{\omega})!}}
|m_1(\bm{\omega}),m_2(\bm{\omega})\rangle
\end{eqnarray}
where we used the notation $(1^{n_1},2^{n_2})$
to indicate the multiplicities of the occurrence of the indices 1 and
2 in the columns of the matrix
$\bm{\Lambda}[\bm{\omega}|(1^{n_1},2^{n_2})]$, as described in
Sec.~\ref{sec:defs}. 
The important object to compute is now the linear entropy. It is
straightforward to see that one can write the trace over the square of
the reduced density matrix as
\begin{eqnarray}
\label{eq:tracerho12}
\mbox{Tr}\hat{\varrho}_1^2 &=& \sum\limits_{k_1,k_2}
\langle k_1,k_2| \hat{U} |\psi_1,\psi_2\rangle \langle \psi_1,\psi_2|
\hat{U}^\dagger |k_1,k_2\rangle
\nonumber \\ &&
\langle \psi_1,\psi_2| \hat{U}^\dagger |k_2\rangle \langle k_2|
\hat{U} |\psi_1,\psi_2\rangle \,.
\end{eqnarray}
This shows that we need all ingredients from the previous section,
that is, both matrix elements of a unitary operator and projections or
partial matrix elements. It is now easy to see that the matrix
elements $\langle k_1,k_2| \hat{U} |\psi_1,\psi_2\rangle$ can be
written as
\begin{eqnarray}
\label{eq:matelem1}
\lefteqn{
\langle k_1,k_2| \hat{U} |\psi_1,\psi_2\rangle =
\sum\limits_{n_1,n_2=0}^N \frac{c_{n_1} d_{n_2}}{\sqrt{k_1!k_2!n_1!n_2!}}
} \nonumber \\ &&
\per{\bm{\Lambda}}[(1^{k_1},2^{k_2})|(1^{n_1},2^{n_2})]
\delta_{n_1+n_2,k_1+k_2} 
\end{eqnarray}
In complete analogy, we can cast the matrix elements
$\langle \psi_1,\psi_2| \hat{U}^\dagger |k_2\rangle$
$\langle k_2| \hat{U} |\psi_1,\psi_2\rangle$ in the form
\begin{eqnarray}
\label{eq:matelem2}
\lefteqn{
\langle \psi_1,\psi_2| \hat{U}^\dagger |k_2\rangle \langle k_2|
\hat{U} |\psi_1,\psi_2\rangle = } \nonumber \\ &&
\sum\limits_{m_1,m_2=0}^N \sum\limits_{n_1,n_2=0}^N
\frac{c_{m_1}^\ast d_{m_2}^\ast c_{n_1} d_{n_2}}
{k_2!(n_1+n_2-k_2)!\sqrt{m_1!m_2!n_1!n_2!}}
\nonumber \\ &&
\per{\bm{\Lambda}}^\ast[(1^{m_1+m_2-k_2},2^{k_2})|(1^{m_1},2^{m_2})]
\nonumber \\ &&
\per{\bm{\Lambda}}[(1^{n_1+n_2-k_2},2^{k_2})|(1^{n_1},2^{n_2})]
\,\delta_{m_1+m_2,n_1+n_2} \,.
\end{eqnarray}
Thus, Eq.~(\ref{eq:tracerho12}) can be written in functional form, by
using Eqs.~(\ref{eq:matelem1}) and (\ref{eq:matelem2}), in terms of
permanents associated with the beam splitter matrix $\bm{\Lambda}$. 

Although the number of sums to be performed to compute the trace of
the square of the reduced density matrix counts up to ten, there are
already three Kronecker $\delta$ functions that reduce the number of
sums to seven. Written out explicitly, Eq.~(\ref{eq:tracerho12}) reads 
\begin{widetext}
\begin{eqnarray}
\mbox{Tr}\hat{\varrho}_1^2 &=& \sum\limits_{k_1,k_2}
\sum\limits_{p_1,p_2=0}^N \sum\limits_{q_1,q_2=0}^N
\sum\limits_{m_1,m_2=0}^N \sum\limits_{n_1,n_2=0}^N
\frac{c_{p_1}d_{p_2}c_{q_1}^\ast d_{q_2}^\ast c_{m_1}^\ast
d_{m_2}^\ast c_{n_1} d_{n_2}}
{k_1!k_2!\sqrt{p_1!p_2!q_1!q_2!}k_2!(n_1+n_2-k_2)!\sqrt{m_1!m_2!n_1!n_2!}} 
\nonumber \\ &&
\per{\bm{\Lambda}}[(1^{k_1},2^{k_2})|(1^{p_1},2^{p_2})]
\per{\bm{\Lambda}}^\ast[(1^{k_1},2^{k_2})|(1^{q_1},2^{q_2})
\nonumber \\ &&
\per{\bm{\Lambda}}^\ast[(1^{m_1+m_2-k_2},2^{k_2})|(1^{m_1},2^{m_2})]
\per{\bm{\Lambda}}[(1^{n_1+n_2-k_2},2^{k_2})|(1^{n_1},2^{n_2})]
\nonumber \\ &&
\delta_{p_1+p_2,k_1+k_2} \delta_{q_1+q_2,k_1+k_2} \delta_{m_1+m_2,n_1+n_2}
\end{eqnarray}
\end{widetext}

At this point we need to have a look at the restrictions the group
integration imposes. The coefficients $c_i$, $i=1,\ldots,N$, are
complex and their squared moduli add up to unity. Hence, they can be
represented by hyperspherical coordinates and some phases as
\begin{equation}
c_i = e^{i\varphi_i} \cos\Theta_{i+1} \prod\limits_{j=1}^{i}
\sin\Theta_j \,,\quad \Theta_{N+1}=0 \,.
\end{equation}
With the additional restriction that $\varphi_0=0$ (because an overall
phase can be absorbed into the definition of the states) we are left
with $2N$ integration variables per initial state. Going back to
Eq.~(\ref{eq:tracerho12}) one notes that the integration has to be
done over four coefficients of the form $c_i^\ast c_j^\ast c_k c_l$
(and similarly for the coefficients of the second initial
state). Averaging over the phase angles then yields
\begin{eqnarray}
\lefteqn{
\frac{1}{(2\pi)^N} \int\limits_0^{2\pi}
\left( \prod\limits_{n=1}^N d\varphi_n \right)
e^{i(\varphi_i+\varphi_j-\varphi_k-\varphi_l)} } \nonumber \\ &&
=(\delta_{ik}\delta_{jl} +\delta_{il}\delta_{jk})
(1-\textstyle\frac{1}{2}\delta_{ij}) \,.
\end{eqnarray}
Since this type of integration has to be performed for both initial
states, there are only few sums remaining in
Eq.~(\ref{eq:tracerho12}). 

The integration over the angles $\Theta_i$ is done by noting that the
coordinates on an $N$-sphere $S^N$ induce a surface integration measure
\begin{equation}
d\mu(S^N) = \left( \prod\limits_{k=1}^{N-1} \sin^{N-k}\Theta_k \right)
\left( \prod\limits_{l=1}^N d\Theta_l \right) \,.
\end{equation}
Therefore, we find that
\begin{eqnarray}
\frac{1}{V}\int d\mu(S^N) |c_i|^4 &=&
\frac{3}{(N+1)(N+3)} \,,\\
\frac{1}{V}\int d\mu(S^N) |c_i|^2|c_j|^2 &=& \frac{1}{(N+1)(N+3)}
\end{eqnarray}
where $V$ is the surface area of the $N$-sphere. Eventually, we are
left with
\begin{equation}
\overline{c_i^\ast c_j^\ast c_k c_l} =
\frac{(\delta_{ik}\delta_{jl}+\delta_{il}\delta_{jk})(1+\frac{1}{2}\delta_{ij})}
{(N+1)(N+3)} \,.
\end{equation}
Inserting this result into Eq.~(\ref{eq:tracerho12}), we finally obtain
something which depends only at most quartically on permanents of the
type
$|\per{\bm{\Lambda}[(1^{m_1+m_2-k_2},2^{k_2})|(1^{m_1},2^{m_2})]}$,
i.e. on absolute values of certain permanents associated with the beam
splitter matrix $\bm{\Lambda}$.

\subsection{Multi-mode networks}

A final remark should be made about multi-mode networks. In
principle, the calculations can be repeated in this case as well and
the matrix elements be expressed in terms of permanents. However,
since it is not entirely clear how to define entanglement measures (or
even monotones) in a multipartite setup, we are not attempting to use
any partial results in these cases. Obviously, once a measure for
multipartite states has been established, it must be expressible in
terms of permanents, as we have seen.

%%%%%%%%%%%%%%%%%%%%%%%%%%%%%%%%%%%%%%%%%%%%%%%%%%%%%%%%%%%%%%%%%%%%%%
\section{Conclusions}
\label{sec:conclusions}

In this article we have shown that calculation of matrix elements of
unitary operators in the Fock basis are intimately connected with the
computation of permanents. This at first surprising result becomes
entirely obvious if one realizes that the (re-)distribution of photons
in a linear-optical network is a purely combinatorial problem. As a
matter of fact, one can think of these networks as being represented
by a bipartite graph whose adjacency matrix has complex entries.

With the knowledge about the relation between matrix elements of
unitary operators and permanents, it becomes immediately obvious
that the computational complexity of calculating functions that depend
on permanents is the same as that for permanents. Examples
for such algorithms that appear in quantum information theory include
maximizations of success probabilities on linear-optical networks
\cite{Scheel03} or averaging procedures involving unitary operators as
in the entanglement power of such networks.

The advantages of relating unitary transformations to permanents are
twofold. On one hand, unitary transformations of multi-mode photon
states can be written in a rather compact form. On the other hand, we
think it may be possible of gaining more insights into the theory of
permanents of unitary matrices from a quantum-mechanical
viewpoint. The example we gave that the permanent of a unitary matrix
takes its value across the unit disk in the complex plane shows that
quantum-mechanical reasoning can simplify the derivation of such
propositions. We believe that this combinatorial view on
linear-optical networks in quantum information processing can
stimulate the emergence of more intricate results.

\acknowledgments
The author likes to thank K.~Audenaert and J.~Eisert for fruitful
discussions and M.B.~Plenio for carefully reading the manuscript.
This work was funded by the UK Engineering and Physical Sciences
Research Council (EPSRC).

%%%%%%%%%%%%%%%%%%%%%%%%%%%%%%%%%%%%%%%%%%%%%%%%%%%%%%%%%%%%%%%%%%%%%%

\end{document}